\documentclass[runningheads]{llncs}
\usepackage{graphicx}
\usepackage{ulem}
\usepackage{tabularx}
\usepackage{booktabs}
\usepackage{caption}
\usepackage{subcaption}
\usepackage{bm}
\usepackage{hyperref}
\usepackage{xcolor}
\usepackage{todonotes}

\begin{document}
\title{How Different are Pre-trained Transformers\\ for Text Ranking?}
%
%
\author{David Rau \and Jaap Kamps}
\authorrunning{D. Rau \and J. Kamps}
%
\institute{University of Amsterdam \\
\email{\{d.m.rau, kamps\}@uva.nl}
}
\maketitle              
\begin{abstract}
In recent years, large pre-trained transformers have led to substantial gains in performance over traditional retrieval models and feedback approaches.  However, these results are primarily based on the MS Marco/\-TREC Deep Learning Track setup, with its very particular setup, and our understanding of why and how these models work better is fragmented at best.
We analyze effective BERT-based cross-encoders versus traditional BM25 ranking
for the passage retrieval task where the largest gains have been observed, and investigate two main questions.  
On the one hand, what is similar? To what extent does the neural ranker already encompass the capacity of traditional rankers? Is the gain in performance due to a better ranking of the same documents (prioritizing precision)?
On the other hand, what is different?  
Can it retrieve effectively documents missed by traditional systems (prioritizing recall)?
We discover substantial differences in the notion of relevance identifying strengths and weaknesses of BERT that may inspire research for future improvement.
Our results contribute to our understanding of (black-box) neural rankers relative to (well-understood) traditional rankers, help understand the particular experimental setting of MS-Marco-based test collections.

\end{abstract}

\section{Introduction}
\label{sec:int}

Neural information retrieval has recently experienced impressive performance gains over traditional term-based methods such as BM25 or Query-Likelihood~\cite{craswell2020overview,craswell2021overview}.   
Nevertheless, its success comes with the caveat of extremely complex models that are hard to interpret and pinpoint their effectiveness. 

With the arrival of large-scale ranking dataset MS MARCO~\cite{bajaj2016ms} massive models such as BERT~\cite{delvinbert} found their successful application in text ranking. Due to the large capacity of BERT (110m+ parameters), it can deal with long-range dependencies and complex sentence structures. When applied to ranking BERT can build deep interactions between query and document that allow uncovering complex relevance patterns that go beyond the simple term matching.
Up to this point, the large performance gains achieved by the BERT Cross-Encoder are not well understood. Little is known about underlying matching principles that BERT bases its estimate of relevance on, what features are encoded in the model, and how the ranking relates to traditional sparse rankers such as BM25 \cite{robertson1994some}. In this work, we focus on the Cross-Encoder (CE) BERT  that captures relevance signals directly between query and document through term interactions between them and refer from now on to the BERT model as \textit{CE}. 
First, we aim to gain a deeper understanding of how CE and BM25 rankings relate to each other, particularly for different levels of relevance by answering the following research questions: 
\begin{itemize}
    \item[] \textbf{RQ1}: How do CE and BM25 rankings vary?
    \item[] \textbf{RQ1.2}: Does CE better rank the same documents retrieved by BM25?
    \item[] \textbf{RQ1.3}: Does CE better find documents missed by BM25?
\end{itemize}

Second, we isolate and quantify the contribution of exact- and soft-term matching to the overall performance. To examine those are particularly interesting as they pose the most direct contrast between the matching paradigms of sparse- and neural retrieval. More concretely, we investigate: 
\begin{itemize}
    \item[] \textbf{RQ2}: Does CE incorporate "exact matching"? 
    \item[] \textbf{RQ3}: Can CE still find "impossible" relevant results?
\end{itemize}
 
\section{Related Work}
\label{sec:rel}
Even though little research has been done to understand the ranking mechanism of BERT previous work exists.
\cite{qiao2019understanding}, \cite{padigela2019investigating}, \cite{zhan2020analysis}, have undertaken initial efforts to open ranking with BERT as a black-box and empirically find evidence that exact term matching and term importance seem to play in an important role. Others have 
tested and defined well-known IR axioms  \cite{camara2020diagnosing}, \cite{rennings2019axiomatic}, \cite{formal2021white} or tried to enforced those axioms through regularization \cite{rosset2019axiomatic}. Another interesting direction is to enforce sparse encoding and able to relate neural ranking to sparse retrieval \cite{zamani2018neural},  \cite{formal2021splade}. 
Although related, the work in \cite{wang2021bert} differs in two important aspects. First, they examine dense BERT retrievers which encode queries and documents independently. Second, they focus rather on the interpolation between BERT and BM25, whereas we specifically aim to understand how the two rankings relate to each other.

\section{Experimental Setup}

The vanilla BERT Cross-Encoder (CE) encodes both queries and documents at the same time. Given input $\bm{x} \in \{[CLS], q_1, \dots, q_n\, [SEP], d_1, \dots , d_m, [SEP]\}$, where $q$ represents query tokens and $d$ document tokens, the activations of the CLS token are fed to a binary classifier layer to classify a passage as relevant or non-relevant; the relevance probability is then used as a relevance score to re-rank the passages. 

We conduct our experiments on the TREC 2020 Deep Learning Track's passage retrieval task on the MS MARCO dataset \cite{bajaj2016ms}. For our experiments, we use the pre-trained model released by \cite{Nogueira2019PassageRW}. 
To obtain the set of top-1000 documents we use anserini's \cite{anserini} BM25 (default parameters) without stemming, following \cite{craswell2020overview}. Table~\ref{tab:bas} shows the baseline performance of BM25 and a vanilla BERT based cross-ranker (CE), re-ranking the 1,000 passages. 

\begin{table*}[!t]
    \centering
    \caption{Performance of BM25 and crossencoder rankers on the NIST judgements of the TREC Deep Learning Task 2020.}
    \label{tab:bas}
\begin{tabularx}{1\textwidth}{l|XXX} 
\toprule 
Ranker \phantom{s} & \phantom{s} NDCG@10& MAP& MRR  \\
\midrule 
BM25 & \phantom{s} 49.59 & 27.47 & 67.06 \\
BERT Cross-Encoder (CE) \phantom{s} & \phantom{s} 69.33& 45.99 & 80.85 \\
\bottomrule
\end{tabularx}
\end{table*}

\section{Experiments}
\label{sec:exp}

\subsection{RQ1: How do CE and BM25 rankings vary?} 
\begin{figure}[!t]
        \centering
        \begin{subfigure}{.5\textwidth}
        \includegraphics[scale=0.35]{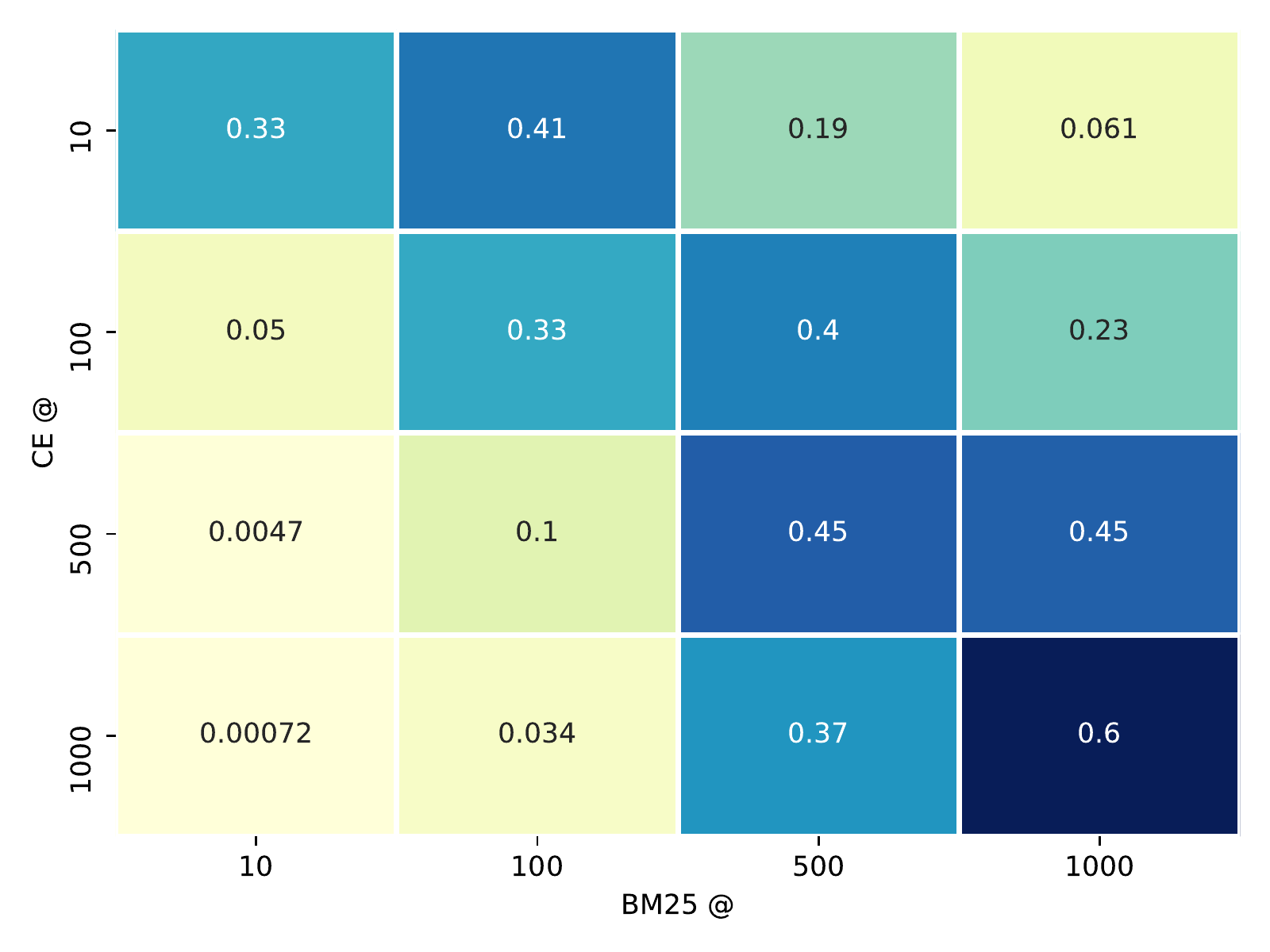}
        \caption{all}
        \end{subfigure}%
         \begin{subfigure}{.5\textwidth}
        \includegraphics[scale=0.35]{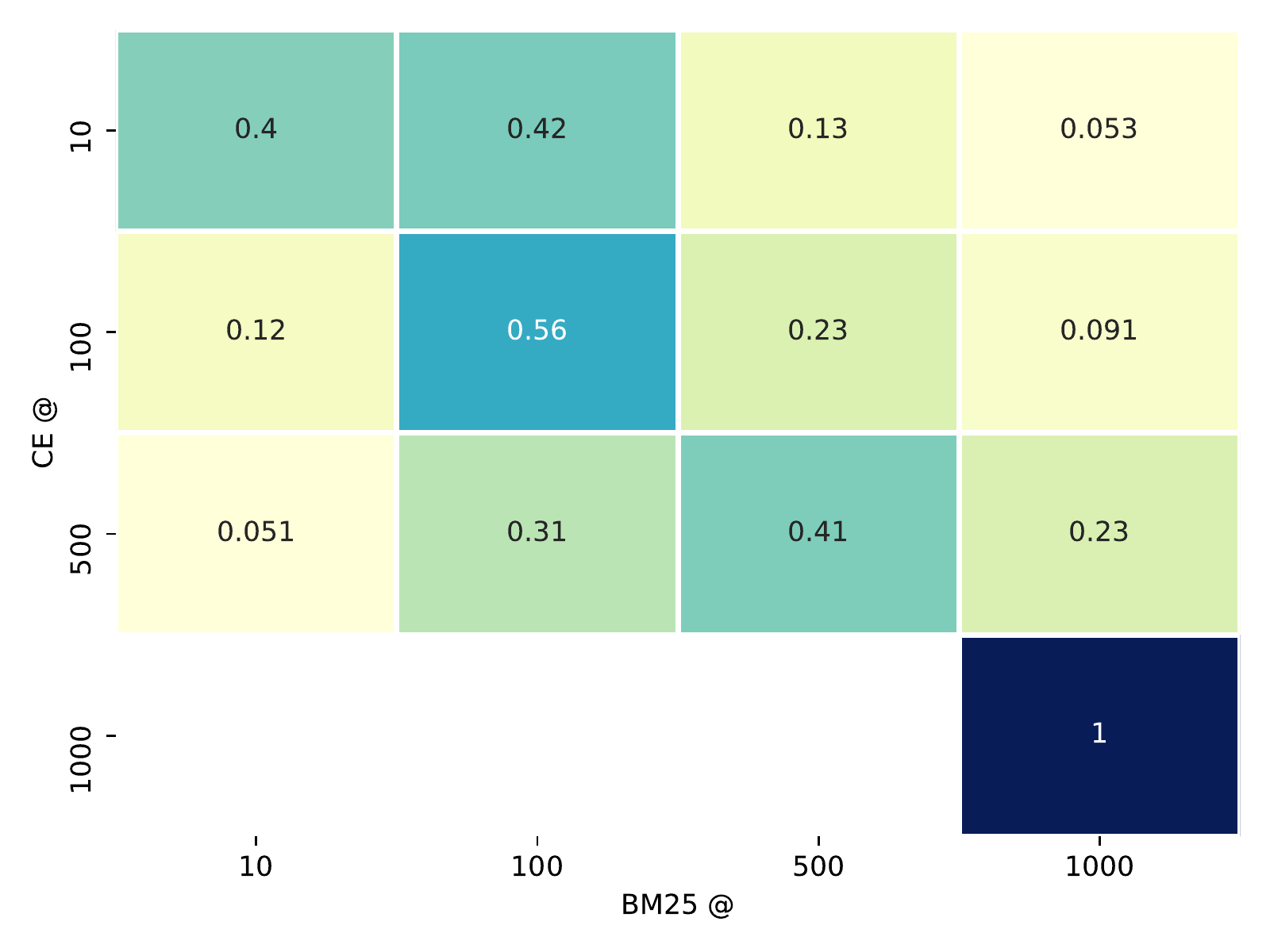}
        \caption{highly relevant}
         \end{subfigure}
        \begin{subfigure}{.5\textwidth}
        \includegraphics[scale=0.35]{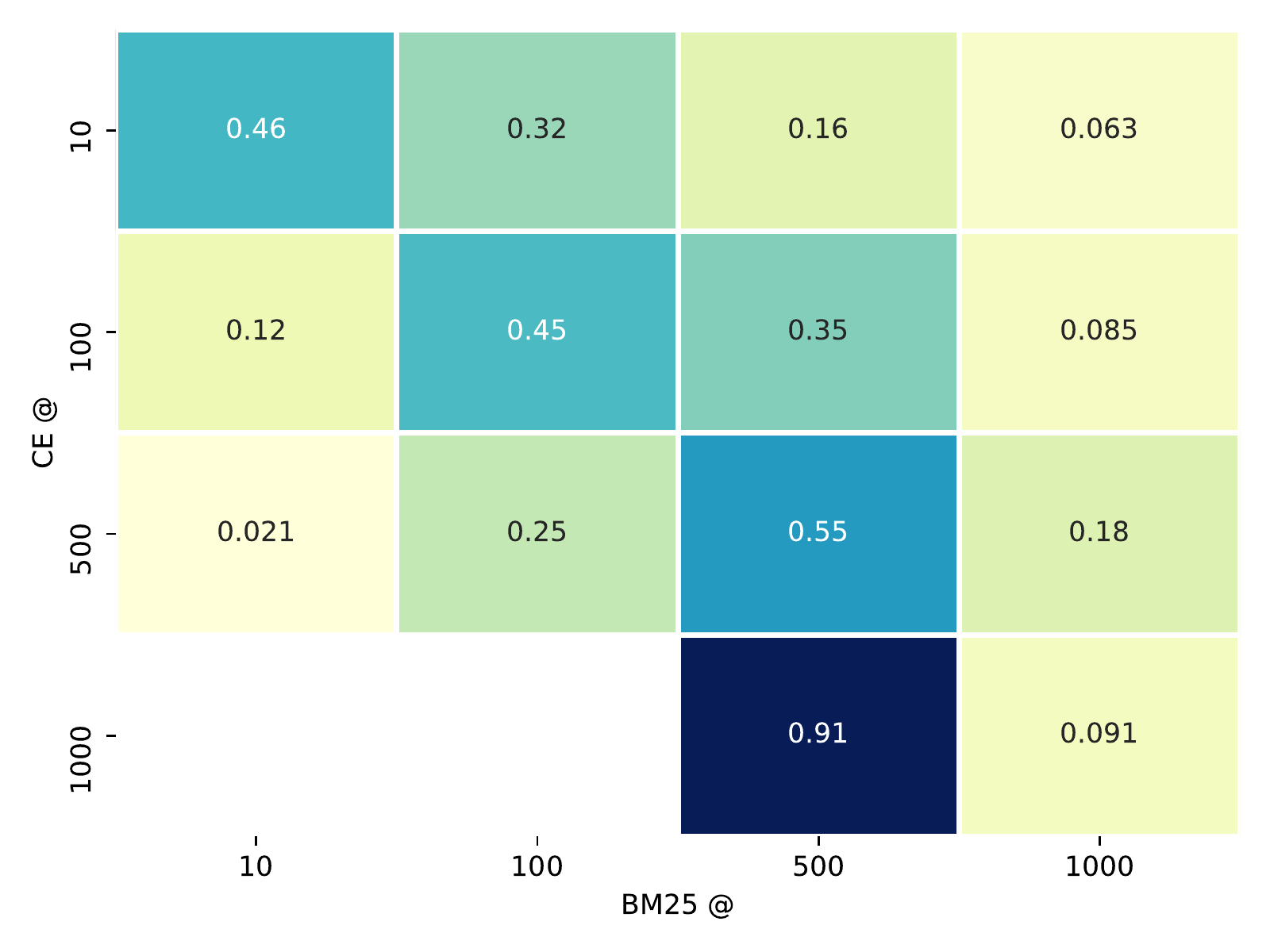}
        \caption{relevant }
        \end{subfigure}%
         \begin{subfigure}{.5\textwidth}
        \includegraphics[scale=0.35]{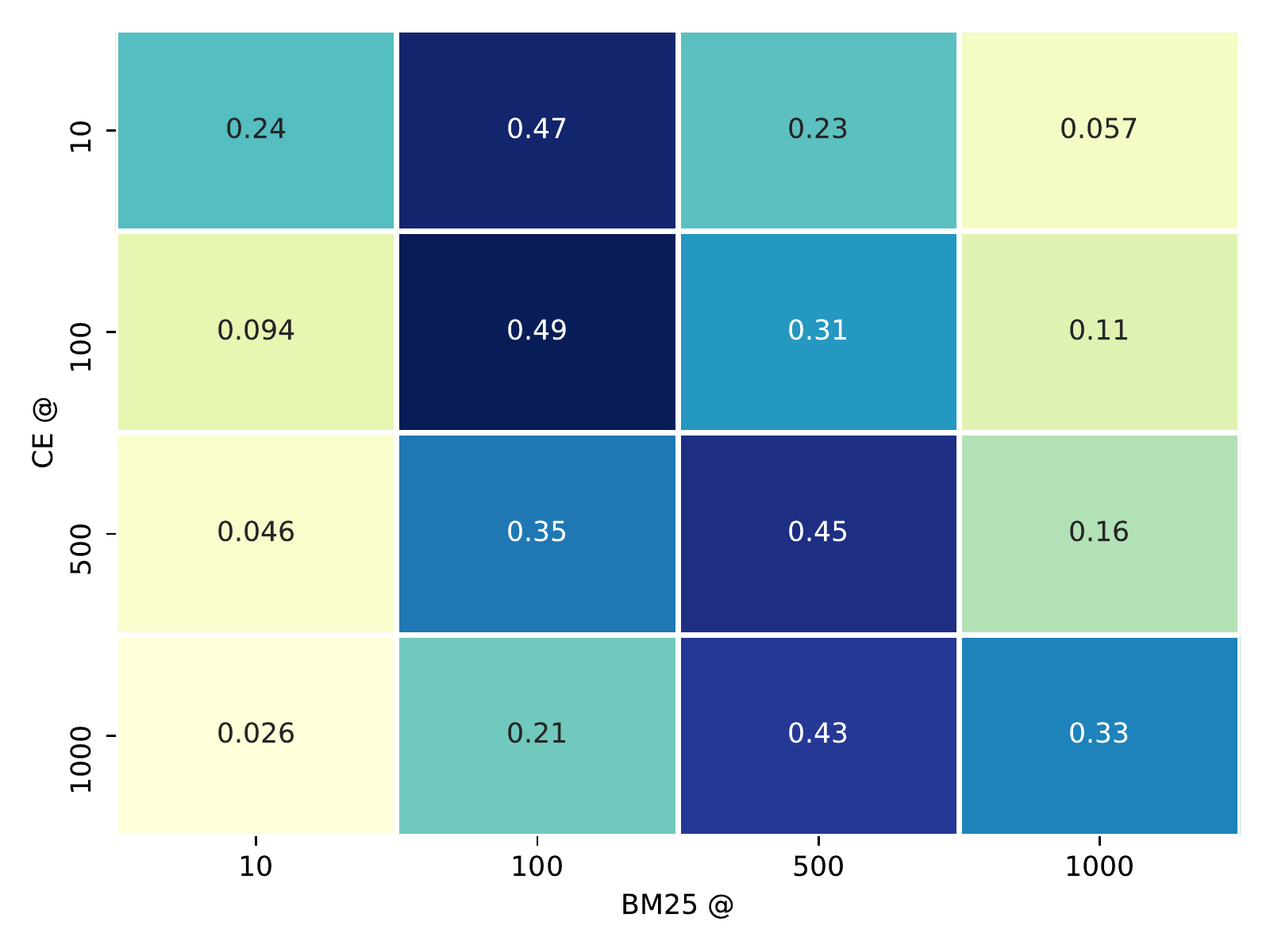}
        \caption{non-relevant}
         \end{subfigure}
     \caption{\small \textbf{Ranking differences between BERT Cross-Encoder (CE) and BM25}: Origin of documents in CE ranking at different rank-ranges with respect to the initial BM25 ranking. More intuitively, each row indicates to what ratio documents stem from different rank-ranges.
     E.g., the top row can be read as the documents in rank 1-10 of the CE re-ranking originate 33\% from rank 1-10, 41\% from rank 11-100, 19\% from rank 101-500 and 6.1\% from rank 501-1000 in the initial BM25 ranking. The rank compositions are shown for (a) all, (b)  highly relevant, (c) relevant, and (d) non-relevant documents according to the NIST 2020 relevant judgments.}
     \label{fig:mov}
 \end{figure}

CE outperforms BM25 by a large margin across all metrics (see Tab. \ref{tab:bas}). To understand the different nature of the CE we trace where documents were initially ranked in the BM25 ranking. For this we split the ranking in different in four rank-ranges: 1-\textit{10}, 11-\textit{100}, 101-\textit{500}, 501-\textit{1000} and will refer to them with ranges \@10, \@100, \@500 and \@1000 respectively from now on. We observe in which rank-range the documents were positioned with respect to the initial BM25 ranking. We show the results in form of heatmaps \footnote{The code for reproducing the heatmaps can be found under \url{https://github.com/davidmrau/transformer-vs-bm25}} in Figure \ref{fig:mov}.

Our initial goal is to obtain general differences between the ranking of CE and BM25 by considering all documents of the test collection (see Fig. \ref{fig:mov} (a)). 
First, we note that CE and BM25 vary substantially on the top of the ranking (33\%  CE@10), whereas at low ranks (60\% CE@1000) the opposite holds. Second, we note that CE is bringing many documents up to higher ranks. Third, we observe that documents ranked high by BM25 are rarely ranked low by CE, suggesting exact matching to be a an important underlying ranking strategy.


\subsection{RQ1.2: Does  CE  better  rank  the  same  documents  retrieved  by  BM25?}
To answer RQ1.2 we consider documents that were judged highly relevant or relevant according to the NIST judgments 2020. The results can be found in Fig. \ref{fig:mov} (b),(c) respectively. Most strikingly, both rankers exhibit a low agreement (40\%) on the documents in CE@10 for \textit{highly relevant} documents hinting a substantial different notion of relevance for the top of the ranking of both methods.

For \textit{ relevant} documents we observe CE and BM25 overlap 46\% at the top of the ranking and a large part (32\%) comes from BM25@100, implying BM25 underestimated the relevance of many documents. The highest agreement between CE and BM25 here is in CE@500 (91\%). 

Interestingly, highly relevant documents that appear in lower ranks originate from high ranks in BM25 (CE@100: 12\%, CE@500: 5\%).
This is an interesting finding as CE fails and underestimates the relevance of those documents, while BM25 - being a much simpler ranker - ranks them correctly. The same effect is also present for \textit{relevant} documents.
When considering documents that both methods ranked low we find a perfect agreement for @1000, showing that the two methods identify the same (highly-)relevant documents as irrelevant.

What about \textit{non-relevant} documents that end up high in the ranking? 
CE brings up to CE@10 a large amount of non-relevant documents from low ranks (47\%  BM25@100, 23\%  BM25@500, and 5\%  BM@1000). Therewith overestimating the relevance of many documents that were correctly considered less relevant by BM25. We also note the little agreement of non-relevant documents @1000 (33\%), hinting at a different notion of irrelevance.

\subsection{RQ1.3: Does  CE  better  find  documents  missed  by  BM25?}
To answer RQ1.3 we again consider documents that were judged (b) highly relevant and (c) relevant and refer to Fig. \ref{fig:mov}, especially focusing on CE@10. The nature of CE, being too expensive for running it on the whole corpus, allows us to only study recall effects within the top-1000 documents. Hence, studying the top-10 results of CE will inform us best about the recall dynamics at high ranks. According to results in Fig. \ref{fig:mov} (b) almost half (42\%) of the highly relevant documents that are missed by BM25 are brought up from BM25@100, 13\% from range BM25@500, and 5\% from range BM25@1000. The same effect can be observed for relevant documents. 
This demonstrates the superior ability of CE to pull up (highly)-relevant documents that are missed by BM25 even from very low ranks. This is the domain where the true potential of the neural models over exact matching techniques lies.

\subsection{RQ2: Does CE incorporate "exact matching"? }
The presence of query words in the document is one of the strongest signals for relevance in ranking~\cite{sara:rele75}, \cite{salt:intr86}. Our goal is to isolate the exact term matching effect, quantify its contribution to the performance, and relate it to sparse ranking. For this, we simply replace all non-query terms in the document with the [MASK] token leaving the model only with a skeleton of the original document and thus forcing it to rely on the exact term matches between query and document only. We do not fine-tune the model on this input. Note that there are no query document pairs within the underlying BM25 top-1000 run that have no term overlap. Results can be found in Tab. \ref{tab:man} under Only Q. CE with only the query words performs significantly lower than BM25 with regard to all metrics, finding clear support that CE is not leveraging exact matches sufficiently.


As in view of finding potential ways to improve CE, our results suggest that exact term matching can be improved.

\begin{table*}[!t]
    \centering
    \caption{Performance of keeping only or removing the query terms from the input.}
    \label{tab:man}
\begin{tabularx}{1\textwidth}{X|XXX}
\toprule 
Model input \phantom{s} & \phantom{s} NDCG@10& MAP& MRR  \\
\midrule
Only Q &  \phantom{s} 31.70 & 18.56 & 44.38 \\
Drop Q & \phantom{s} 49.89 & 29.08  & 65.12 \\
\bottomrule
\end{tabularx}
\end{table*}

\subsection{RQ3: Can CE still find "impossible" relevant results?}

While CE can leverage both, exact term- as well as "soft" matches, the biggest advantage over traditional sparse retrievers holds the ability to overcome lexical matches and to take context into account. Through "soft" matches neural models can retrieve documents that are "impossible" to retrieve using traditional potentially resulting in high recall gains. To isolate and quantify the effect of "soft matches" we follow our previous experiment but this time mask the appearance of the query words in the document. The model has now to rely on the surrounding context only. We do not fine-tune the model on this input. Note that in this setting BM25 would score randomly. Results can be found in Tab. \ref{tab:man} under Drop Q.
We observe that CE can score documents sensibly with no overlapping query terms, largely outperforming when ranking on query terms only (Only Q). The model scores 49.89 NDCG@10 points losing only around 20 points with respect to non-manipulated input. CE might be able to fill-in the masked tokens from the context, as this makes up a main part of the Masked-Language modeling pre-training task. The model demonstrates its true potential here by drawing on its ability to understand semantics through the contextualization of query and document and to leverage its associate memory.

    
\section{Conclusions and Discussion}
\label{con}
Our experiments find evidence that documents at the top of the ranking are generally ranked very differently while a stronger agreement at the bottom of the ranking seems to be present.
By investigating the rankings for different relevance levels we gain further insight. Even though, for (highly-)relevant documents there exists a bigger consensus at the top of the ranking compared to the bottom we find a discrepancy in the notion of high relevance between them for some documents, highlighting core differences between the two rankers. 

We discover that CE is dramatically underestimating some of the highly relevant documents that are correctly ranked by BM25. This sheds light on the sub-optimal ranking dynamics of CE, sparking clues to overcome current issues to improve ranking in the future. 
Our analysis finds further evidence that the main gain in precision stems from bringing (highly-)relevant documents up from lower ranks (early precision). On the other hand, CE overestimates the relevance of many non-relevant documents where BM25 scored them correctly lower.

Through masking all but the query words within the documents we show that CE is not able to rank on the basis of only exact term matches only scoring a lot  lower than BM25. By masking the query words in the document we demonstrate the ability of CE to score queries and documents without any lexical overlap with a moderate loss of performance, therefore demonstrating the true strength of neural models over traditional methods, that would completely fail in this scenario, in isolation.

We leave it to further research to qualitatively investigate the query-document pairs that BERT fails, but BM25 ranks correctly. 

%
%
%
\subsubsection*{Acknowledgments} 
{\small
This research is funded in part by 
the Netherlands Organization for Scientific Research 
(NWO CI \# CISC.CC.016), and
the Innovation Exchange Amsterdam (POC grant).
}
\bibliographystyle{splncs04}
\bibliography{bibl}
\end{document}